\input{aipcheck}

\documentclass[
    ,final            
  ]
  {aipproc}

\layoutstyle{6x9}

\begin{document}

\title{Double beta decay to the excited states: experimental review}


\author{A.S. Barabash}{
  address={Institute of Theoretical and Experimental Physics, 
B. Cheremushkinskaya 25, 117218 Moscow, Russia}
}



\begin{abstract}
 A brief review on double beta decay to excited states of 
daughter nuclei is given. The ECEC($0\nu$) transitions to the excited states 
are discussed in association with a possible enhancement of the 
decay rate by several orders of magnitude.
\end{abstract}

\maketitle


\section{Introduction}

  The $\beta\beta$ decay can proceed through transitions to the ground 
state as well as to various excited states of the daughter nucleus. 
Studies of the latter transitions allow 
supplementary information about $\beta\beta$ decay.
The first experimental studies of $\beta\beta$ 
decay to the excited state was done by E. Fiorini in 1977 \cite{FIO77}.
It was just an aside to his main experiment 
with $^{76}$Ge (transition to 0$^+$ ground state). First special experimental work to investigate 
the $\beta\beta$ decay to the excited states were done in 1982 \cite{BEL82}.
In 1989 it was shown that using low-background facilities utilizing 
High Purity Germanium (HPGe) detectors, the $2\nu\beta\beta$ decay to the 0$^+_1$
level in the daughter nucleus may be detected for such nuclei as $^{100}$Mo, 
$^{96}$Zr and $^{150}$Nd \cite{BAR90}. Soon after double beta decay of $^{100}$Mo to 
the 0$^+$ excited state at 1130.29 keV in $^{100}$Ru was observed 
\cite{BAR95}. Then this result
was confirmed in a few independent experiments with HPGe detectors \cite{BAR99,DEB01,HOR06}. 
In 2004 for the first 
time this transition was detected in $^{150}$Nd 
\cite{BAR04}. Recently the $2\nu\beta\beta$ decay of $^{100}$Mo to the 0$^+_1$
level in $^{100}$Ru was detected using tracking detector NEMO-3 where all the 
decay products (two electrons and two $\gamma$-rays) were detected
and hence all the information about the decay was obtained (total 
energy spectrum, single electron spectrum, single $\gamma$ spectrum and all 
angular distributions) \cite{ARN07}.  In addition in the last 15 years new  
limits for many nuclei and different modes of decay to the excited states 
were established (see reviews \cite{BAR00,BAR04a}). 
Present motivations to do this search are the following:

1) Nuclear spectroscopy (to know decay schemes of nuclei).

2) Nuclear Matrix Elements.

3) Examination of some new ideas (such as the "bosonic" component of the neutrino,
 \cite{DOL05,BAR07}).

4) Neutrino mass investigations: 
  
 a) $0\nu\beta\beta(0^+-0^+_1)$ decay; in this case one has a very nice signature 
for the decay and hence high sensitivity to neutrino mass could be reached; 
   
b) high sensitivity to the effective Majorana neutrino mass can be reached in the case of 
the ECEC (0$\nu$) transition
if the resonance condition is realized (see \cite{BER83,SUJ04,BAR07a}).

\section{Double beta decay to the excited states}

The present experimental status of $\beta\beta$ decay 
to the excited states of daughter nuclei is the following.

\subsection{$2\nu\beta\beta$ transition to the $2^+_1$ excited state}

The $2\nu\beta\beta$ decay to the $2^+_1$ excited state is strongly suppressed 
and practically inaccessible to detection. However, for a few nuclei ($^{96}$Zr, 
$^{100}$Mo, $^{130}$Te) there are some "optimistic" predictions for half-lives 
($T_{1/2}$ $\sim$ $10^{22}-10^{24}$ y)
and there is a chance to detect such decays in the next generation of 
the double beta decay experiments.
The best present limits are shown in Table 1.

\begin{table}
\begin{tabular}{lrrrr}
\hline
  
   \tablehead{1}{r}{b}{Nuclei}
  & \tablehead{1}{r}{b}{E$_{2\beta}$,\\keV}
  & \tablehead{1}{r}{b}{Experiment\\$T_{1/2}$, y}
  & \tablehead{1}{r}{b}{Theory\\\cite{RAD07}}
  & \tablehead{1}{r}{b}{Theory\\\cite{SUH96,SUH97}}   \\
\hline
$^{48}$Ca & 3288.5 & $> 1.8\times10^{20}$ \cite{BAK02} & $1.7\times10^{24}$ & -\\
$^{150}$Nd  & 3033.6 & $> 9.1\times10^{19}$ \cite{ARP99}  & -  & - \\
$^{96}$Zr  & 2572.2 & $> 7.9\times10^{19}$ \cite{BAR96} & $2.3\times10^{25}$ & $(3.8-4.8)\times10^{21}$ \\
$^{100}$Mo  & 2494.5 & $> 1.6\times10^{21}$ \cite{BAR95} & $1.2\times10^{25}$ & $3.4\times10^{22}$ \cite{STO96} \\
$^{82}$Se  & 2218.5 & $> 1.4\times10^{21}$ \cite{SUH97a} & - & $2.8\times10^{23}$-$3.3\times10^{26}$ \\
$^{130}$Te  & 1992.7 & $> 2.8\times10^{21}$ \cite{BEL87} & $6.9\times10^{26}$ & $(3.0-27)\times10^{22}$  \\
$^{116}$Cd  & 1511.5 & $> 2.3\times10^{21}$ \cite{PIE94} & $3.4\times10^{26}$ & $1.1\times10^{24}$ \\
$^{76}$Ge  & 1480 & $> 1.1\times10^{21}$ \cite{BAR95a} & $5.8\times10^{28}$ & $(7.8-10)\times10^{25}$  \\

\hline
\end{tabular}
\caption{Best present limits on $2\nu\beta\beta$ transition to the $2^+_1$ excited state (90\% C.L.). }
\label{tab:a}
\end{table}

\subsection{$2\nu\beta\beta$ transition to the $0^+_1$ excited state}

For these transitions the best results and limits are presented in Table 2. 
Table 3 presents all the 
existing positive results for $2\nu\beta\beta$ decay of $^{100}$Mo to the first $0^+$
excited state of $^{100}$Ru. The half-life averaged over all
four experiments is given in the bottom row. The average value was calculated using the 
standard procedure of determining the average for different accuracy measurements , 
and the statistical and systematic errors were summed quadratically (for more 
details see \cite{BAR06}).

\begin{table}
\begin{tabular}{lrrrr}
\hline
  
   \tablehead{1}{r}{b}{Nuclei}
  & \tablehead{1}{r}{b}{E$_{2\beta}$,\\keV}
  & \tablehead{1}{r}{b}{Experiment\\$T_{1/2}$, y}
  & \tablehead{1}{r}{b}{Theory\\\cite{SUH96,SUH97}}
  & \tablehead{1}{r}{b}{Theory\\\cite{STO96}}   \\
\hline
$^{150}$Nd & 2627.1 & $= 1.4^{+0.5}_{-0.4}\times10^{20}$ \cite{BAR04} & - & -\\
$^{96}$Zr  & 2202.5 & $> 6.8\times10^{19}$ \cite{BAR96}  & $(2.4-2.7)\times10^{21}$    & $3.8\times10^{21}$  \\
$^{100}$Mo  & 1903.7 & $= 6.2^{+0.9}_{-0.7}\times10^{20}$ & $1.6\times10^{21}$ \cite{HIR95} & 
$2.1\times10^{21}$ \\
$^{82}$Se  & 1507.5 & $> 3.0\times10^{21}$ \cite{SUH97a} & $(1.5-3.3)\times10^{21}$ & - \\
$^{48}$Ca  & 1274.8 & $> 1.5\times10^{20}$ \cite{BAK02} & - & - \\
$^{116}$Cd  & 1048.2 & $> 2.0\times10^{21}$ \cite{PIE94} & $1.1\times10^{22}$ & $1.1\times10^{21}$  \\
$^{76}$Ge  & 916.7 & $> 6.2\times10^{21}$ \cite{KLI02} & $(7.5-310)\times10^{21}$ & $4.5\times10^{21}$ \\
$^{130}$Te  & 735.3 & $> 2.3\times10^{21}$ \cite{BAR01} & $(5.1-14)\times10^{22 *)}$ & - \\

\hline
\end{tabular}
\caption{Best present results and limits on $2\nu\beta\beta$ transition to the $0^+_1$ excited state. 
Limits are given at the 90\% C.L. $^{*)}$ Corrected value is used (see \cite{BAR01}).}
\label{tab:a}
\end{table}

\begin{table}
\begin{tabular}{lrrrr}
\hline
  
   \tablehead{1}{r}{b}{$T_{1/2}$, y}
  & \tablehead{1}{r}{b}{N}
  & \tablehead{1}{r}{b}{S/B}
  & \tablehead{1}{r}{b}{Year, References}
  & \tablehead{1}{r}{b}{Method}   \\
\hline
$ 6.1^{+1.8}_{-1.1}\times10^{20}$  & 66 & $\sim 1/7$ & 1995 \cite{BAR95} & HPGe\\
$9.3^{+2.8}_{-1.7} \pm 1.4\times 10^{20}$  & 80 & $\sim 1/4$ & 1999 \cite{BAR99}& HPGe \\
$6.0^{+1.9}_{-1.1} \pm 0.6\times 10^{20}$ & 19.5 & 8/1 & 2001 \cite{DEB01,HOR06} & 2xHPGe \\
$5.7^{+1.3}_{-0.9} \pm 0.8\times 10^{20}$ & 37.5 & 3/1 & 2007 \cite{ARN07} & NEMO-3 \\
\hline
Average value: $ 6.2^{+0.9}_{-0.7}\times10^{20}$ y  \\
\hline
\end{tabular}
\caption{Present "positive" results on $2\nu\beta\beta$ decay of $^{100}$Mo to the first $0^+$
excited state of $^{100}$Ru. N is the number of useful events, S/B is the signal-to-background ratio.}
\label{tab:a}
\end{table}

\subsection{$0\nu\beta\beta$ transition to the $2^+_1$ excited state}

The $0\nu\beta\beta (0^+-2^+_1)$ decay had long been accepted to be possible because 
of the contribution of right-handed currents and is not sensitive to the neutrino mass contribution.
However, in Ref. \cite{TOM00} it was demonstrated that the relative sensitivities of ($0^+-2^+_1)$ decays
to the neutrino mass $\langle m_{\nu} \rangle$ and the right-handed current $\langle \eta \rangle$  
are comparable to those of $0\nu\beta\beta$ decay to the ground state. At the same time, the
($0^+-2^+_1)$ decay is more sensitive to $\langle \lambda \rangle$.     
The best present experimental limits are shown in Table 4.

\begin{table}
\begin{tabular}{lrrrr}
\hline
  
   \tablehead{1}{r}{b}{Nuclei}
  & \tablehead{1}{r}{b}{E$_{2\beta}$,\\keV}
  & \tablehead{1}{r}{b}{Experiment\\$T_{1/2}$, y}
  & \tablehead{1}{r}{b}{Theory \cite{TOM00}, \\$\langle m_{\nu} \rangle$ = 1 eV }
  & \tablehead{1}{r}{b}{Theory \cite{TOM00}, \\ $\langle \lambda \rangle$ = $10^{-6}$ }   \\
\hline
$^{76}$Ge & 1480 & $> 8.2\times10^{23}$ \cite{MAI94} & $8.2\times10^{31}$ & $6.5\times10^{29}$  \\
$^{100}$Mo & 2494.5 & $> 1.6\times10^{23}$ \cite{ARN07} & $6.8\times10^{30}$ & $2.1\times10^{27}$ \\
$^{130}$Te  & 1992.7 & $> 1.4\times10^{23}$ \cite{ARN03} & - & - \\
$^{116}$Cd  & 1511.5 & $> 2.9\times10^{22}$ \cite{DAN03} & - & - \\
$^{136}$Xe  & 1649.4 & $> 6.5\times10^{21}$ \cite{BEL91}  & - & - \\
$^{82}$Se  & 2218.5 & $> 2.8\times10^{21}$ \cite{ARN98} & - & -  \\

\hline
\end{tabular}
\caption{Best present limits on $0\nu\beta\beta$ transition to the $2^+_1$ excited state (90\% C.L.).}
\label{tab:a}
\end{table}

\subsection{$0\nu\beta\beta$ transition to the $0^+_1$ excited state}

The $0\nu\beta\beta$ transition to the $0^+$ excited states of the daughter nuclei provides 
a clear-cut signature. In addition to two 
electrons with a fixed total energy, there are two photons, whose energies are strictly 
fixed as well. In a hypothetical experiment detecting all decay products with high efficiency 
and high energy resolution, the background can be reduced to nearly zero. It is possibl this idea
will be used in future experiments featuring a large mass of the isotope under study (as 
mentioned in Refs. \cite{BAR04a,BAR00,SUH01}). In Ref. \cite{SIM02} it was mentioned that 
detection of this transition will 
give us the additional possibility to distinguish the $0\nu\beta\beta$ mechanisms. 
The best present limits are presented in Table 5.

\begin{table}
\begin{tabular}{lrrrr}
\hline
  
   \tablehead{1}{r}{b}{Nuclei}
  & \tablehead{1}{r}{b}{E$_{2\beta}$,\\keV}
  & \tablehead{1}{r}{b}{Experiment\\$T_{1/2}$, y}
  & \tablehead{1}{r}{b}{Theory\\\cite{SUH00,SUH01,SUH02,SUH03}}
  & \tablehead{1}{r}{b}{Theory\\\cite{SIM01}}   \\
\hline
$^{150}$Nd & 2627.1 & $> 1.0\times10^{20}$ \cite{ARP99} & - & -\\
$^{96}$Zr  & 2202.5 & $> 6.8\times10^{19}$ \cite{BAR96}  & $2.4\times10^{24}$    & -  \\
$^{100}$Mo & 1903.7 & $> 8.9\times10^{22}$ \cite{ARN07} & $2.6\times10^{26}$ & $1.5\times10^{25}$ \\
$^{82}$Se  & 1507.5 & $> 3.0\times10^{21}$ \cite{SUH97a} & $9.5\times10^{26}$ & $4.5\times10^{25}$   \\
$^{48}$Ca  & 1274.8 & $> 1.5\times10^{20}$ \cite{BAK02} & - & - \\
$^{116}$Cd  & 1048.2 & $> 1.4\times10^{22}$ \cite{DAN03} & $1.5\times10^{27}$ & -  \\
$^{76}$Ge  & 916.7 & $> 1.3\times10^{22}$ \cite{MOR88} & $4.9\times10^{26}$ & $2.4\times10^{26}$ \\
$^{130}$Te  & 735.3 & $> 3.1\times10^{22}$ \cite{ARN03}  & $7.5\times10^{25}$ & - \\

\hline
\end{tabular}
\caption{Best present limits on $0\nu\beta\beta$ transition to the $0^+_1$ excited state (90\% C.L.). 
Theoretical predictions 
for $\langle m_{\nu} \rangle$ = 1 eV are given. }
\label{tab:a}
\end{table}

\section{ECEC(0$\nu$) transition to the excited states}

In Ref. \cite{WIN55} it was the first mentioned that in the case of ECEC(0$\nu$)
transition a resonance condition could exist for transition to a "right energy" excited level 
of the daughter nucleus (when decay energy is closed to zero).
In 1982 the same idea was proposed for transition to the ground state \cite{VOL82}. 
In 1983 this possibility was discussed for the transition $^{112}$Sn-$^{112}$Cd ($0^+$; 1871 keV)  
\cite{BER83}. In 2004 the idea 
was reanalyzed in Ref. \cite{SUJ04} and some new resonance condition for the decay was formulated. 
The possible enhancement of the transition rate was estimated as $\sim$ 10$^6$ \cite{BER83,SUJ04}. 
This means that this process starts to be
competitive with $0\nu\beta\beta$ decay for the sensitivity to neutrino mass
and it is interesting to check this idea by experiment.
There are several candidate for such resonance transition, to the ground ($^{152}$Gd, $^{164}$Eu and 
$^{180}$W) and to the excited states ($^{74}$Se, $^{78}$Kr, $^{96}$Ru, $^{106}$Cd, $^{112}$Sn, 
$^{130}$Ba, $^{136}$Ce and $^{162}$Er)) of daughter nuclei (see \cite{BAR07,SUJ04}).
The precision needed to realize resonance condition is well below 1 keV. To select the best 
candidate from the above list one will have to know the atomic mass difference with an 
accuracy better then 1 keV.
Such measurements are planed for the future. Recently the first experiment to search for such
a resonance transition in $^{74}$Se-$^{74}$Ge ($2^+$; 1206.9 keV) was perfomed yielding a 
limit $T_{1/2} > 5.5\times10^{18}$
y \cite{BAR07a}. It was also demonstrated that using enriched $^{74}$Se and an installation 
such as GERDA or MAJORANA
a sensitivity on the level $\sim 10^{26}$ y can be reached.




\bibliographystyle{aipproc}   




\end{document}